\documentclass[pss,fleqn,embeddedheads]{w-art}
\usepackage{times}
\usepackage{w-thm}
\usepackage[]{graphicx}
\begin{document}
\DOIsuffix{theDOIsuffix}
\Volume{XX}
\Issue{1}
\Copyrightissue{01}
\Month{01}
\Year{2003}
\pagespan{1}{}
\Receiveddate{\sf zzz} \Reviseddate{\sf zzz} \Accepteddate{\sf
zzz} \Dateposted{\sf zzz}




\title[Charge Transfer and Charge Transport on the Double Helix]
{Charge Transfer and Charge Transport on the Double Helix}


\author[N.P. Armitage]{N.P. Armitage\footnote{Corresponding
     author: e-mail: {\sf npa@physics.ucla.edu}}}
\address{Department of Physics and Astronomy,
University of California, Los Angeles, CA 90095} 

\author[M. Briman]{M. Briman}

\author[G. Gr\"{u}ner]{G. Gr\"{u}ner}

\begin{abstract}

We present a short review of various experiments that measure
charge transfer and charge transport in DNA.  Some general
comments are made on the possible connection between
\textit{'chemistry-style'} charge transfer experiments that probe
fluorescence quenching and remote oxidative damage and
\textit{'physics-style'} measurements that measure transport
properties as defined typically in the solid-state. We then
describe measurements performed by our group on the millimeter
wave response of DNA. By measuring over a wide range of humidity
conditions and comparing the response of single strand DNA and
double strand DNA, we show that the appreciable AC conductivity of
DNA is not due to photon assisted hopping between localized
states, but instead due to dissipation from dipole motion in the
surrounding water helix.
\end{abstract}

\maketitle                   




\renewcommand{\leftmark}
{N.P. Armitage $et$ $al.$: Charge Transfer and Charge Transport on
the Double Helix}

\section{Overview}

The electrical conductivity of DNA has been a topic of much recent
interest and controversy \cite{Dekker}. Measurements from
different groups have reached a variety of conclusions about the
nature of charge transfer and transport along the double helix.
Although there has been a flurry of recent activity, the subject
has long history. Eley and Spivey in 1962 \cite{EleySpivey} were
the first to note that the unique structure of DNA with $\pi-\pi$
orbital stacking separated by 3.4 $\AA$ resembled high mobility
aromatic crystals and suggested it as efficient structure for
electron transfer.

Charge transfer is one of the most fundamental chemical processes,
driving such disparate reactions as corrosion and photosynthesis.
 The semi-classical Marcus \cite{marcus} theory predicts an
exponential charge transfer efficiency that falls off as
$e^{-\beta r}$ with $\beta \approx 1.5$ $\AA$.  These
considerations seemed borne out by two decades of experiments on
proteins and other $\sigma$-bonded network bridges between
photoexcited metal complexes and electron acceptors. Hence,
initial experiments \cite{Bartoninitial} probing the $\pi$-bond
stack of DNA that showed the possibility of longer range charge
transfer were surprising. In these first experiments, fluorescent
molecules bound to calf thymus DNA were quenched by the addition
of electron acceptors to the strands. They suggested a transfer
efficiency $e^{-\beta r}$ with $\beta \approx 0.2$ $\AA$. The
expectations of Eley and Spivey not-withstanding this was counter
to the prevailing paradigm of transfer efficiency $\beta \approx
1.5$ $\AA$ from the Marcus theory. Such long range mobile
electrons raised the possibility of interesting electronic effects
on the double helix. Transfer along this supposed '$\pi$-way' was
referred to as 'wire-like'. This work prompted many other
experiments to be done, both within the chemistry community and
within the solid-state physics community the latter attempting to
measure the transport properties of DNA directly. The activity has
lead to new theories, such as polaron transport \cite{Kawai} and
conformational gating \cite{Bruinsma}, regarding charge transfer
and transport in molecular stacks and biological systems .

Additional experiments showed that the value of $\beta$ obtained
seemed to depend on the details of the strand sequences and
donor-acceptor complex used. In the initial experiments Murphy
$et$ $al.$ \cite{Bartoninitial} tethered a ruthenium intercalator
to end of a single DNA strand and a rhodium intercalator to a
complementary stand. When annealed, ruthenium luminescence was
completely quenched by the rhodium intercalator positioned almost
40 $\AA$ down the $\pi$-stack. With the organic intercalator
ethidium \cite{Kelley} as the photoexcited donor and rhodium as
the acceptor similar quenching behavior was shown over distances
of 20 to 30 $\AA$. However other organic donor-acceptor complexes
showed $\beta \approx 1$ $\AA$ \cite{Brun}. Lewis $et$
$al.$\cite{Lewis}, using stilbene as fluorescence at the end of an
A-T chain, systematically moved a G-C pair (functioning as an
acceptor) away from the stilbene. They found that quenching rate
decreased quickly until about 4 separating A-T's and then more
slowly after that.

So-called 'chemistry-at-a-distance' by electron transfer was shown
by radical induced strand cleavage.  Meggers $et$ $al.$
\cite{Meggers} formed a highly oxidizing radical guanine cation at
one end of a DNA strand that had a GGG unit on the other end. The
GGG unit is purported to have a lower ionization potential than a
single G and hence can accept the hole which neutralizes the
radical G. The strand was then treated to cleave at the resulting
oxidation site. The length dependence of the electron transfer
could be found by varying the number of intervening bridge states
and performing electrophoresis to find the number and lengths of
cleaved strands. The measurements showed exquisite sensitivity to
intervening T-A bases. The efficiency was found to be determined
by the longest 'bottleneck' – i.e. the longest hopping T-A step.

Strong evidence that the charge transfer was truly happening
through base-base hopping via the $\pi-\pi$ overlap was given by
measurements that probed changes in oxidized guanine damage yield
with response to base perturbations
\cite{stackpertub1,stackpertub2}. Overall the efficacy of charge
transfer through the mismatch was found to correlate with how well
bases in the mismatch were stacked.  This gave strong evidence
that charges are transferred through the $\pi-\pi$ stack directly.

These measurements, taken as a whole, gave an emerging picture
where a hole has its lowest energy on the G­C sites and for short
distances moves from one G­C pair to the next by coherent
tunnelling through the A­T sites. The overall motion from the
initial base pair to the last is an incoherent hopping mechanism
i.e. the charged carrier is localized on sites along the path. For
longer distances between G-C base pairs the picture was that
thermal hopping onto A-T bridges becomes the dominant
charge-transfer mechanism which gave the weaker distance
dependence above four separating A-T pairs of Lewis et al.
\cite{Lewis}. Under such circumstances $\beta$ becomes a poor
parametrization of the transfer efficiency as the distance
dependence is no longer exponential. Such a picture has been
supported from the quantum-mechanical computation models of Burin,
Berlin, and Ratner \cite{Ratner}.

These 'chemistry-style' experiments give convincing evidence that
electron or holes can delocalize over a number of base pairs and
that the extent of the delocalization is governed by strand
sequence among other aspects. Although such experiments have
motivated the direct measure of transport properties via DC and AC
techniques, the information gained from luminescence quenching
measurements and the like is not directly related to their
conductivity i.e. the ability to behave as a molecular wire.
Although the descriptor 'wire-like' has been applied to sequences
where a small $\beta$ has been found, such terminology is
misleading.

Luminesce quenching is an excited state property.  Under
appreciated by the solid state community working in this field is
the relatively large energy scale ($\approx2 eV$) of the typical
redox potentials for a luminence quenching reaction
(stilbene*/stilbene: 1.75 eV and Rh-complex+3/Rh-complex+2: 2 eV).
 In solid-state physics jargon these are very high energy
electron-hole excitations.  Perhaps a good solid-state analog of
this phenomenon is the luminescence quenching of fluorescent atoms
doped into semiconductors, as for instance in Si:Er or ZnSe:Cu
\cite{SiEr1,SiEr2,ZnCeCu}.  In erbium doped silicon a photoexcited
electron-hole pair is captured by an impurity level on Er. Decay
of this level imparts energy to the 4f$Er^{+3}$ system, which then
decays and emits a fluorescent photon. Such a fluorescence can be
quenched by detrapping of the captured electron pair on the Er
level into the conduction band. In this case, the detrapping is
into an orbital which is completely invisible to DC transport.
Such an experiment tells us only that there is finite overlap
between the localized level and some further extended states.  We
learn nothing directly applicable to the material's ability to
conduct electricity. Likewise luminescence quenching in DNA can be
viewed as the detrapping of a hole into the HOMO orbital,
whereupon even weak orbital overlap allows it to make its way to
the acceptor under the influence of the driving redox potential.
Charge transfer experiments confirm delocalization of hole over a
few bases, but we learn little about materials ability to behave
as wire.  In this regard these charge transfer experiments merely
reflect the strong effects of disorder in 1-D i.e. the
localization of all states. A small $\beta$ is not synonymous with
'wire-like' behavior.

Although there is relative agreement among chemists regarding the
charge transfer properties of DNA, the physics community has not
reached a similar d\'{e}tente with respect to measurements of its
direct charge transport properties. DNA has been reported to be
metallic \cite{Fink}, semiconducting \cite{Porath}, insulating
\cite{Braun,dePablo}, and even a proximity effect induced
superconductor\cite{Kasumov}. However, questions have been raised
in many papers with regards to length effects, the role played by
electrical contacts, and the manner in which electrostatic damage,
mechanical deformation by substrate-molecule interaction, and
residual salt concentrations and other contaminants may have
affected these results.  Some recent measurements, where care was
taken to both establish a direct chemical bond between
$\lambda$-DNA and Au electrodes and also control the excess ion
concentration, have given compelling evidence that the DC
resistivity of the DNA double helix over long length scales
($<10\mu$m) is very high indeed ($\rho > 10^6 \Omega - cm$)
\cite{Zhang}.  These results were consistent with earlier work
that found flat I-V characteristics and vanishingly small
conductances \cite{dePablo}, but contrast with other studies that
found a substantial DC conductance that was interpreted in terms
of small polaron hopping \cite{Kawai}.  DC measurements that show
DNA to be a good insulator are also in apparent contradiction with
recent contactless AC measurements that have shown appreciable
conductivity at microwave and far-infrared frequencies
\cite{Tran,Helgren} the magnitude of which approaches that of a
well-doped semiconductor \cite{HelgrenSiP}.

In previous finite frequency studies, the AC conductivity in DNA
was found to be well parameterized as a power-law in $\omega$
\cite{Tran,Helgren}. Such a dependence can be a general hallmark
of AC conductivity in disordered systems with photon assisted
hopping between random localized states \cite{ES} and led to the
reasonable interpretation that intrinsic disorder, counterion
fluctuations, and possibly other sources created a small number of
electronic states on the base pair sequences in which charge
conduction could occur. However, such a scenario would lead to
thermally activated hopping conduction between these localized
states and is thus inconsistent with a very low DC
conductivity\cite{Zhang}.  To the end of resolving some of these
matters, we have extended our previous AC conductivity experiments
in the millimeter wave range to a wide range of humidity
conditions. We show that the appreciable AC conductivity of DNA in
the microwave and far infrared regime should not be viewed as some
sort of hopping between localized states and is instead likely due
to dissipation in the dipole response of the water molecules in
the surrounding hydration layer.

\section{Experimental Details}

Double stranded DNA films were obtained by vacuum drying of 7mM
PBS solution containing 20 mg/ml sodium salt DNA extracted from
calf thymus and salmon testes (Sigma D1501 and D1626).  In order
to improve the DNA/salt mass ratio we used a high concentration of
DNA, but it was found that the limit was 20 mg/ml.  Higher
concentrations makes it difficult for DNA fibers to dissolve and
the solution becomes too viscous, which prevents producing the
flat uniform films which are of paramount importance for the
quasi-optical resonant technique. It was found that as long as the
excess salt mass fraction is kept between 2-5\% the final results
were not significantly affected.  Single stranded DNA films were
prepared from the same original solution as the double stranded
ones.  The solution was heated to 95 C for 30 minutes and the
quickly cooled to 4 C.  We checked the conformational state of
both double-strand DNA(dsDNA) and single-strand DNA (ssDNA) by
fluorescent microscope measurements. Films, when dry, were 20 to
30 microns thick and were made on top of 1mm thick sapphire
windows. Immediately after solution deposition onto the sapphire
substrates the air inside the viscous solution was expelled by
vacuum centrifuging at 500g, otherwise the evaporation process
causes the formation of air bubbles that destroy the film
uniformity.

The AC conductivity was measured in the millimeter spectral range.
Backward wave oscillators (BWO) in a quasi-optical setup (100 GHz
- 1 THz) were employed as coherent sources in a transmission
configuration.  This range, although difficult to access
experimentally,  is particularly relevant as it corresponds to the
approximate expected time frame for relaxation processes in room
temperature liquids (0.1-10 ps). Importantly, it is also below the
energy range where one expects to have appreciable structural
excitations.  The technique and analysis are well established
\cite{Schwartz}.

\section{Results}

We measured samples at room temperature at several fixed humidity
levels.  They were maintained in a hermetically sealed environment
with a saturated salt solution \cite{Falk} that kept moisture
levels constant. The mass of the DNA films and changes in
thickness were tracked by separate measurements within a
controlled environment for each sample in a glove box. The total
number of water molecules per nucleotide $A$ can be correlated to
the relative humidity $x$ ($x=0-1$) through the so-called
Branauer-Emmett-Teller (BET) equation \cite{BET}

\begin{equation}
\centering
 A=\frac{BCx}{(1-x)(1-x+Cx)}.
\end{equation}

The constant $B$ denotes the maximum number of water molecules in
the first layer sites.   Mobile water molecules within the double
helix can be characterized as 2 types according to the statistical
formulation of the BET equation by Hill \cite{Hill}. The first are
those within the initial hydration layer, which are directly
attached to DNA and have a characteristic binding energy
$\epsilon_1$. Water molecules of the second and all other layers
can be approximated as having a binding energy $\epsilon_L$.  To a
good approximation this $\epsilon_L$ can be taken to be that of
bulk water.  These parameters enter into the BET equation through
the expression for $C$ which equals
$De^{(\frac{\epsilon_1-\epsilon_L}{kT})}$ where $D$ is related to
the partition function of water.  Also we should note that there
is, in actuality, a structural 0-th layer of water molecules,
containing 2.5-3 water molecules per nucleotide that cannot be
removed from the helix under typical conditions ~\cite{Tao}.

\begin{vchfigure}[h]
\centering
\includegraphics[width=.4\textwidth]{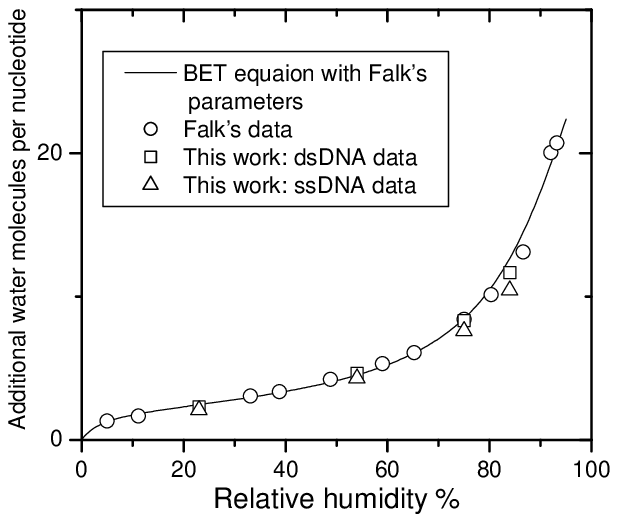}
\vchcaption{Adsorbtion of water molecules per nucleotide as a
function of humidity.  The data represented by the open circles is
taken from Falk \textit{et. al.} [28].} \label{hydration}
\end{vchfigure}

Falk \textit{et al.}'s ~\cite{Falk} first established that the
adsorption of mobile water layers of DNA can be modelled by
distinguishing 2 different types of water parameters by use of the
BET equation to describe the hydration of sodium and lithium DNA
salts. They found good agreement between experimental data and
theory with constants $B=2.2$ and $C=20$. We performed a similar
hydration study of our dsDNA and ssDNA films; as shown in Fig.
\ref{hydration} the hydration of our films is perfectly consistent
with the results of Falk.  We found no appreciable difference in
the hydration between dsDNA and ssDNA.

In Fig. \ref{conduct} data is presented for the extracted
$\sigma_1(\omega)$ of both dsDNA and ssDNA thin films. In both
cases, the conductivity is an increasing function of frequency.
Since the conductivity also increases with humidity, one may wish
to try to separate the relative contributions of charge motion
along the DNA backbone from that of the surrounding water
molecules.

First, one can consider that there should be two main effects of
hydration in our dsDNA films. There is the hydration itself, where
water molecules are added in layers. Additionally, the
conformational state of dsDNA changes as a function of adsorbed
water.  Although water molecules can certainly contribute to the
increase in conductivity,  at high humidities there is the
possibility that some of the conduction might be due to an
increase in electron transfer along the dsDNA helix in the ordered
B form. However since such an effect would be much reduced in
disordered and denaturalized single strand DNA and since Fig. 2
shows that to within the experimental uncertainty the conductivity
of dsDNA and ssDNA in the millimeter wave range is
indistinguishable, it is most natural to suggest that water is the
major contribution to the AC conductivity.  From this comparison
of dsDNA and ssDNA, we find no evidence for charge conduction
along the DNA between bases.

\begin{vchfigure}[htb]
\centering
\includegraphics[width=.5\textwidth]{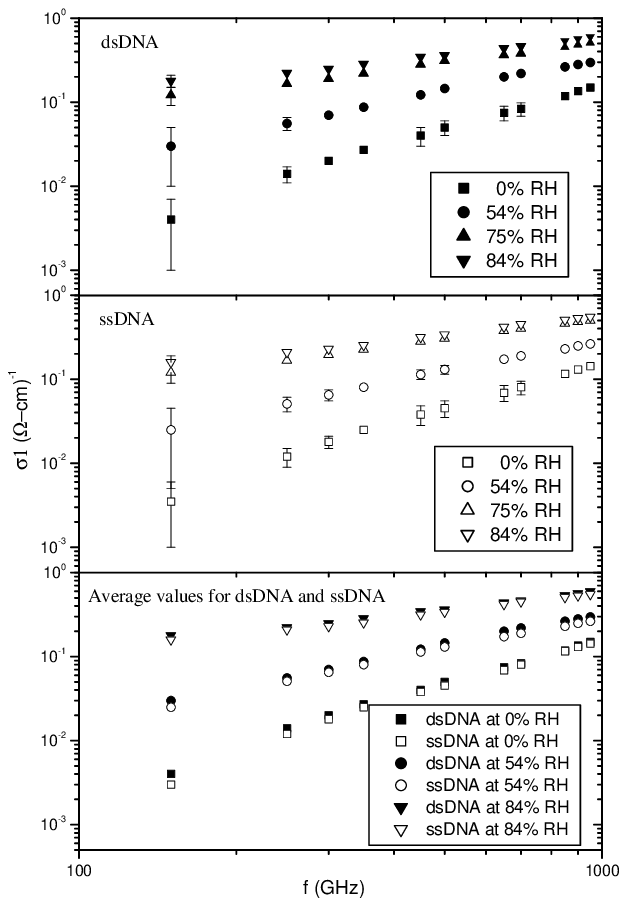}
\vchcaption{AC conductivity of calf thymus DNA at different
relative humidity levels.  (a)  Double stranded DNA (b)  Single
stranded DNA  (c) A comparison of conductivity between single and
double stranded DNA.} \label{conduct}
\end{vchfigure}

\section{Discussion}

In Fig. 3 we plot the conductivity $\sigma_1$ of the DNA films
normalized by the expected volume fraction of water molecules
including both the hydration layers plus the structural water.
Although this normalization reduces the spread in the thin film
conductivity at the lowest frequencies it does not reduce it to
zero, showing that if the observed conductivity comes from water,
the character of its contribution changes as a function of
humidity.

The complex dielectric constant of bulk water has been shown to be
well described by a biexponential Debye relaxation model
\cite{Ronne,Kindt,Barthel}, where the first relaxation process
\cite{Ronne}, characterized by a time scale $\tau_D=8.5$ ps,
corresponds to the collective motion of tetrahedral water
clusters, and the second from faster single molecular rotations
\cite{Agmon} with a time scale $\tau_F=170$ fs. For bulk water,
the contribution of each relaxation process is determined by the
static dielectric constant $\epsilon_S(T) \approx 80$, $
\epsilon_1=5.2$, and the dielectric constant at high frequencies
$\epsilon_{\infty}=3.3$.

\begin{equation}
\widehat{\epsilon}(w)=\epsilon_{\infty}+\frac{\epsilon_S-\epsilon_1}{1+i\omega\tau_D}+
\frac{\epsilon_1-\epsilon_{\infty}}{1+i\omega\tau_F}
\end{equation}

Eq. 2 gives us insight into the conduction and loss processes
occurring in the water layers.  For high hydration levels, where
multiple water layers exist around the dipole helix, the
relaxation losses may approach those of bulk water.  The above
equation can be compared, using the independently known values
\cite{Ronne} for $\tau_D$, $\epsilon_S$, $\tau_F$ and
$\epsilon_1$, to the experimental data normalized to the expected
volume fraction of the water.  The conductivity of well hydrated
DNA is seen to approach that of bulk water.

One expects that the contribution to the loss of cluster
relaxation processes to decrease as the number of water layers
decreases.  As the structural water is not tetrahedrally
coordinated, it is reasonable that first term of Eq. 2, which is
due to the collective motion of water clusters, cannot contribute
at low humidity.  Remarkably, the 0\% humidity conductivity
appears to be described by a model that only includes the fast
single molecule rotation of bulk water.  This is notable because
such behavior is at odds with many systems that find longer net
relaxation times in thin adsorbed gas layers than in the
corresponding bulk systems \cite{Singh}.

In Fig. 3, along with the experimental data at two representative
humidity levels, two theoretical curves for 0\% and 100\% humidity
are plotted. With the only two assumptions being that at 0\%
humidity, the sole relaxational losses come from singly
coordinated water molecules in the structural water layer and that
it is only at higher humidity levels where the collective losses
can gradually play a greater role, the theoretical curves provide
a very good bound to the data over almost all of the measured
frequency range.

\begin{vchfigure}[htb]
\centering
\includegraphics[width=.55\textwidth]{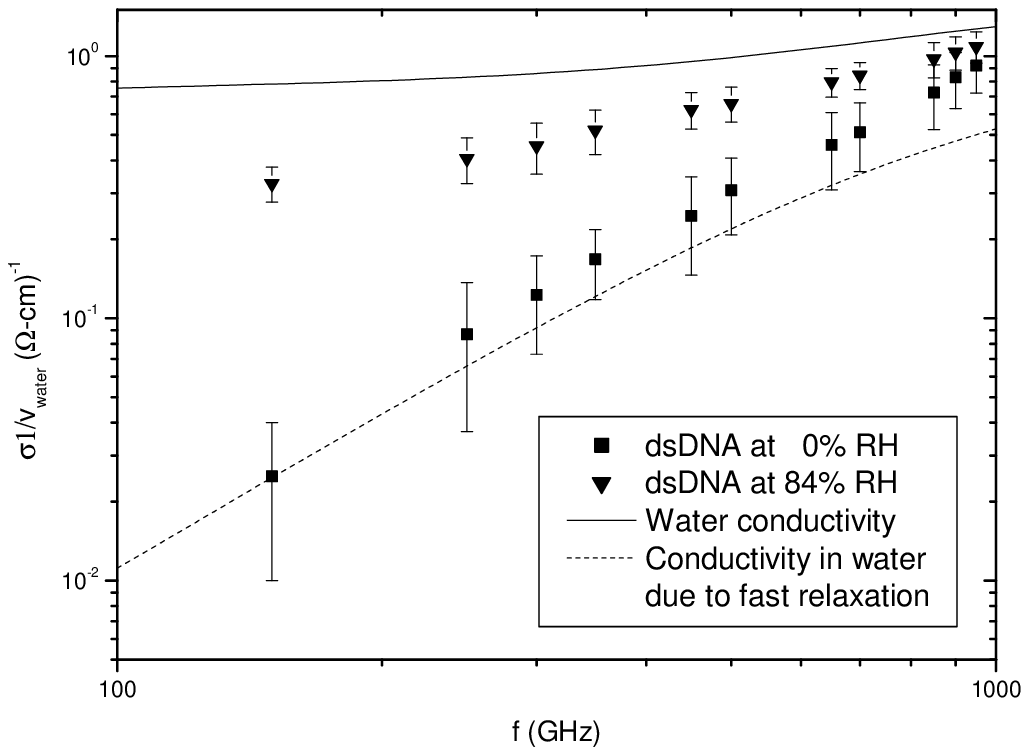}
\vchcaption{Conductivity of dsDNA and ssDNA films normalized by
the volume fraction of all water molecules (structural plus
hydration layer). For clarity, only 0\% and 84\% humidities are
shown.  The solid line represents the conductivity of pure water
as modelled by the biexponential Debye model using the parameters
of Ronne $et$ $al.$ The dashed line shows just the contribution
from single water molecule relaxation.} \label{adjcond}
\end{vchfigure}

The only large difference between the experiment and theory is the
high frequency data at low humidity, where the model
underestimates the conductivity.  There are a number of
possibilities for these discrepancies.  It may be that at higher
frequencies for low hydration samples, the weak restoring force
from charge-dipole interaction in the structural water layer
becomes more significant and our biexponential Debye model is less
applicable.  Alternatively, it is possible that at very low
relative humidities for the ionic phosphate groups on the DNA
backbone to form stable dihydrates which may give their own
contribution to relaxation losses through their additional degree
of freedom \cite{Falk}.  We should also note that one advantage of
working in the millimeter spectral range is the known weak
contribution of ionic conduction in this regime \cite{Jackson}.
The motion of the surrounding relatively large mass counterions
only becomes appreciable at lower frequencies \cite{Slovenia}

\section{Conclusion}

In conclusion, we have found that the considerable AC conductivity
of DNA can be attributed largely to relaxational losses of the
surrounding water dipoles.  The AC conductivity of ssDNA and dsDNA
was found to be identical to within the experimental error.  As
this changes the base-base orbital overlap significantly, this
indicates the absence of charge conduction along the DNA backbone
itself. The conclusion that the observed conductivity derives from
the water layer is supported by the fact that, over much of the
range, it can be well described by a biexponential Debye model,
where the only free parameter is the relative contributions of
single water molecule and tetrahedral water cluster relaxation
modes.

\begin{acknowledgement}
 We would like to thank K. Greskoviak for help with sample
preparation and E. Helgren for assistance at various stages of the
measurements. The research at UCLA was supported by the National
Science Foundation grant DMR-0077251.

\end{acknowledgement}

\end{document}